\documentclass[12pt]{article}
\usepackage{amssymb} 
\usepackage{amsmath}

\begin{document}
\font\frak=eufm10 scaled\magstep1                 
\font\fak=eufm10 scaled\magstep2
\font\fk=eufm10 scaled\magstep3
\font\black=msbm10 scaled\magstep1
\font\bigblack=msbm10 scaled\magstep 2
\font\bbigblack=msbm10 scaled\magstep3
\font\scriptfrak=eufm10
\font\tenfrak=eufm10
\font\tenblack=msbm10


\def\biggoth #1{\hbox{{\fak #1}}}
\def\bbiggoth #1{\hbox{{\fk #1}}}
\def\sp #1{{{\cal #1}}}
\def\goth #1{\hbox{{\frak #1}}}
\def\scriptgoth #1{\hbox{{\scriptfrak #1}}}
\def\smallgoth #1{\hbox{{\tenfrak #1}}}
\def\smallfield #1{\hbox{{\tenblack #1}}}
\def\field #1{\hbox{{\black #1}}}
\def\bigfield #1{\hbox{{\bigblack #1}}}
\def\bbigfield #1{\hbox{{\bbigblack #1}}}
\def\Bbb #1{\hbox{{\black #1}}}
\def\v #1{\vert #1\vert}             
\def\ord#1{\vert #1\vert} 
\def\m #1 #2{(-1)^{{\v #1} {\v #2}}} 
\def\lie #1{{\sp L_{\!#1}}}               
\def\pd#1#2{\frac{\partial#1}{\partial#2}}
\def\pois#1#2{\{#1,#2\}}
\def\set#1{\{\,#1\,\}}             
\def\<#1>{\langle#1\rangle}        
\def\>#1{{\bf #1}}                
\def\f(#1,#2){\frac{#1}{#2}}
\def\cociente #1#2{\frac{#1}{#2}}
\def\braket#1#2{\langle#1\mathbin\vert#2\rangle} 
\def\brakt#1#2{\langle#1\mathbin,#2\rangle}           
\def\dd#1{\frac{\partial}{\partial#1}} 
\def\bra #1{{\langle #1 |}}
\def\ket #1{{| #1 \rangle }}
\def\ddt#1{\frac{d #1}{dt}}
\def\dt2#1{\frac{d^2 #1}{dt^2}}
\def\matriz#1#2{\left( \begin{array}{#1} #2 \end{array}\right) }
\def\Eq#1{{\begin{equation} #1 \end{equation}}}

\def\bw{{\bigwedge}}      
\def\hut{{\scriptstyle \wedge}}            
\def\dg{{\goth g^*}}                                                                                                            
\def\Cdg{{C^\infty (\goth g^*)}}
\def\poi{\{\:,\}}                           
\def\qw{\hat\omega}                
\def\FL{{\sp F}L}                 
\def\hFL{\widehat{{\sp F}L}}      
\def\XHMw{\goth X_H(M,\omega)} 
\def\XLHMw{\goth X_{LH}(M,\omega)}                  
\def\ea{\varepsilon_a}
\def\ep{\varepsilon}
\def\mitad{\frac{1}{2}}
\def\x{\times}  
\def\cinf{C^\infty} 
\def\forms{\bigwedge}                 
\def\onda{\tilde}
\def\orb{{\sp O}}

\def\a{\alpha}
\def\g{{\gamma }}                  
\def\G{{\Gamma}}
\def\La{\Lambda}                   
\def\la{\lambda}                   
\def\w{\omega}                     
\def\W{{\Omega}}                   
\def\ltimes{\bowtie} 
             
\def\roc{{\tilde{\cal R}}}                       
\def\cl{{\cal L}}                               
\def\V{{\sp V}}                                 
\def\F{{\sp F}}
\def\cv{{{\goth X}}}                    
\def\LG{\goth g}
\def\X{{{\goth X}}}                     
\def\R{{\hbox{{\field R}}}}             
\def\big R{{\hbox{{\bigfield R}}}}
\def\bbig R{{\hbox{{\bbigfield R}}}}
\def\C{{\hbox{{\field C}}}}         
\def\Z{{\hbox{{\field Z}}}}             
\def\N{{\hbox{{\field N}}}}         

\def\ima{\hbox{{\rm Im}}}                               
\def\dim{\hbox{{\rm dim}}}        
\def\End{\hbox{{\rm End}}} 
\def\Tr{\hbox{{\rm Tr}}} 
\def\tr{{\hbox{\rm\small{Tr}}}}                
\def\lin{{\hbox{Lin}}}
\def\vol{{\hbox{vol}}}  
\def\Hom{{\hbox{Hom}}}
\def\div{{\hbox{div}}} 
\def\rank{{\hbox{rank}}}
\def\Ad{{\hbox{Ad}}}
\def\ad{{\hbox{ad}}}
\def\CoAd{{\hbox{CoAd}}}
\def\coad{{\hbox{coad}}}                           
\def\Rea{\hbox{Re}}                     
\def\id{{\hbox{id}}}                    
\def\Id{{\hbox{Id}}}
\def\Int{{\hbox{Int}}}
\def\Ext{{\hbox{Ext}}}
\def\Aut{{\hbox{Aut}}}
\def\Card{{\hbox{Card}}}
\def\SODE{{\small{SODE }}}


\newtheorem{teor}{Teorema}[section]
\newtheorem{cor}{Corolario}[section]
\newtheorem{prop}{Proposici\'on}[section]
\newtheorem{definicion}{Definici\'on}[section]
\newtheorem{lema}{Lema}[section]

\newtheorem{theorem}{Theorem}
\newtheorem{corollary}{Corollary}
\newtheorem{proposition}{Proposition}
\newtheorem{definition}{Definition}
\newtheorem{lemma}{Lemma}

\def\Eq#1{{\begin{equation} #1 \end{equation}}}
\def\R{\Bbb R}
\def\C{\Bbb C}
\def\Z{\Bbb Z}
\def\d{\partial}

\def\la#1{\lambda_{#1}}
\def\teet#1#2{\theta [\eta _{#1}] (#2)}
\def\tede#1{\theta [\delta](#1)}
\def\N{{\frak N}}
\def\Wei{\wp}
\def\Hil{{\cal H}}

\font\frak=eufm10 scaled\magstep1

\def\bra#1{\langle#1|}
\def\ket#1{|#1\rangle}
\def\goth #1{\hbox{{\frak #1}}}
\def\<#1>{\langle#1\rangle}
\def\cotg{\mathop{\rm cotg}\nolimits}
\def\wt{\widetilde}
\def\const{\hbox{const}}
\def\grad{\mathop{\rm grad}\nolimits}
\def\Div{\mathop{\rm div}\nolimits}
\def\braket#1#2{\langle#1|#2\rangle}
\def\Erf{\mathop{\rm Erf}\nolimits}

\centerline{\Large \bf A new geometric approach to Lie systems} 

\bigskip

\centerline{\Large \bf and physical applications}

\vskip 2cm

\centerline{\sc Jos\'e F. Cari\~nena and Arturo Ramos}

\vskip 0.5cm

\centerline{Departamento de  F\'{\i}sica Te\'orica, Universidad de Zaragoza,}
\medskip
\centerline{50009 Zaragoza, Spain.}
\medskip
\centerline{email: {\tt jfc@posta.unizar.es} and {\tt arrg@posta.unizar.es}}

\vskip 1cm

\begin{abstract}   
The characterization of systems of differential equations 
admitting a superposition function allowing us to write 
the general solution in terms of any fundamental set of 
particular solutions is discussed. These systems are
shown to be related with equations on a Lie group and with some 
connections in fiber bundles. We develop two methods for dealing 
with such systems: the generalized Wei--Norman method and the
reduction method, which is very useful when particular solutions
of the original problem are known. The theory is illustrated with some
applications in both classical and quantum mechanics.
\end{abstract}

\section{Introduction} 
Time evolution of many physical systems is  described  by non-autonomous 
systems of 
differential equations
\begin{equation}
\frac{dx^i(t)}{dt} = X^i(t,x)\ , \qquad i=1,\ldots,n\,,
\label{tdynsyst}
\end{equation}
 for instance,  Hamilton equations or Lagrange 
equations when transformed to the first order case by 
doubling the number of degrees of freedom.

The Theorem of existence and uniqueness of solution 
 for such systems establishes that
 the initial conditions $x(0)$ determine the future evolution. It is also
 well-known that for the simpler case of a homogeneous linear system the 
general solution can be written as a linear combination of $n$ independent 
particular solutions, $x_{(1)}, \ldots ,x_{(n)}$,
 \begin{equation}
x=F(x_{(1)},\ldots,x_{(n)},k_1,\ldots,k_n)
=k_{1}\, x_{(1)}+\cdots +k_{n}\,x_{(n)}\ ,\label{lsr} 
\end{equation}
and for each set of initial conditions, 
 the coefficients can be determined. For
an inhomogeneous linear system, the general solution can be written
 as an affine function of $(n+1)$ independent particular solutions:
\begin{eqnarray}
&&x=F(x_{(1)},\ldots,x_{(n+1)},k_1,\ldots,k_n) \nonumber\\
&&\quad\quad\quad=x_{(1)}+ k_1(x_{(2)}-x_{(1)})+\ldots+k_n(x_{(n+1)}-x_{(1)})\ . \label{asr}  
\end{eqnarray}
Under a non-linear change of coordinates both system become  
non-linear ones. However, the fact that the general solution is expressible 
in terms of a set of particular solutions is maintained, 
but the superposition function is no longer linear or affine, respectively.

The very existence of such examples of systems of differential equations 
admitting a superposition function suggests us an analysis of such systems.
We are lead in this way to the problem of characterizing the systems 
of differential equations for which a superposition function, 
allowing to express the general solution in terms of $m$ particular solutions, 
does exist. The solution of this problem is due to Lie \cite{LS}.
Our aim here is to review the theory developed by Lie from a modern 
geometric viewpoint and to present different applications both
in mathematics and physics.

The paper is organized as follows: Section 2 present the main Theorem 
due to Lie and some simple examples are given in Section 3. In Section 4,
after the 
introduction of some notation concerning ingredients of Lie group theory, the
particular case in which the  systems are defined in a Lie group $G$
is analyzed,
and we show how to relate them with a particular type of equations on a group.
We also show that Lie systems in homogeneous spaces arise naturally associated
 with these systems in Lie groups. The theory is illustrated with a pair 
of examples which  point out the universal character of the equation in the
 group.  Section 5 is devoted to present a generalization to the general case 
of a method 
proposed by Wei and Norman for linear systems and the
 example of the affine group in one dimension is used to illustrate the 
theory. The relation of the problem at hand with the theory of connections
 is studied in Section 6: it 
is shown that Lie systems define horizontal curves with respect to a 
connection. The reduction method developed in section 7 corresponds to 
consider the action of the group of automorphisms of the principal bundle 
on the set of connections, transforming in this way the given problem in a
 simpler one. Some examples and references to different applications of
 this reduction method
are also given. The applications of the general theory to different problems in 
both Classical and Quantum Mechanics are indicated in Section 8 with an 
especial  emphasis on time evolution of time-dependent Hamiltonian systems.
The example of the time-dependent linear potential model  has been explicitly developed  in Section 9 in both the classical and the quantum case.

\section{Lie Theorem}

The characterization of non-autonomous systems  (\ref{tdynsyst}) having the 
mentioned property that the general solution can be written as a function of $m$
independent particular solutions and some constants determining each specific
solution is due to Lie. The statement of the Theorem, which can be found in the 
book edited and revised by Scheffers \cite{LS}, is as follows:

\begin{theorem}
{Given a non-autonomous system of  $n$ first order
differential equations like (\ref{tdynsyst}),
a necessary and sufficient condition for the existence of a function 
$F:{\Bbb R}^{n(m+1)}\to {\Bbb R}^n$ such that the general solution is
$$x=F(x_{(1)}, \ldots,x_{(m)};k_1,\ldots,k_n)\ ,$$ 
with $\{x_{(a)}\mid a=1,\ldots,m\}$
being any set of particular solutions of the system
and  $k_1,\ldots,k_n,$ 
 $n$  arbitrary constants,
is that the system can be written as  
\begin{equation}
 \frac {dx^i}{dt}=Z_1(t)\xi^{1i}(x)+\cdots+Z_r(t)\xi^{ri}(x)\,,\qquad i=1,
\ldots,n\,,  
\end{equation}
where  $Z_1,\ldots,Z_r,$ are  $\,r$ functions  depending only on  $t$ and
  $\xi^{\alpha i}$, $\alpha=1,\ldots, r$,  are functions of 
 $x=(x^1,\ldots,x^n)$,
such that the  $r$ vector fields in   ${\Bbb R}^n$ given by
\begin{equation}
Y^{(\alpha)}\equiv   \sum_{i=1}^n\xi^{\alpha i}(x^1,\ldots,x^n)
\pd {}{x^i}\,,\qquad
 \alpha=1,\ldots,r,
\end{equation}
close on a real  finite-dimensional Lie algebra, i.e. the vector fields 
$Y^{(\alpha)}$ are linearly independent  
and  there exist $r^3$
real numbers, $f^{\alpha\beta}\,_\gamma$, such that
\begin{equation}
[Y^{(\alpha)},Y^{(\beta)}]=\sum_{\gamma=1}^rf^{\alpha\beta}\,_\gamma Y^{(\gamma)}\ .
\end{equation}
}
\label{Lie_Theorem}
\end{theorem}
The number  $r$ satisfies $r\leq m\,n$. For a geometric proof, see \cite{CGM}.

{}From the geometric viewpoint, the system of first order differential 
equations  (\ref{tdynsyst}) provides the integral curves of the $t$-dependent 
vector field on an $n$-dimensional manifold $M$
$$X =  \sum_{i=1}^nX^i(x,t)\pd{}{x^i}\ ,$$
 in the same way as it happens for autonomous systems and true vector fields, 
 and the $t$-dependent vector fields satisfying the hypothesis of the
 Theorem are those which
can be written as a $t$-dependent linear combination of  vector fields, 
$$X(x,t)=\sum_{\alpha=1}^r Z_\alpha(t)\, Y^{(\alpha)}(x)\ ,$$ with
vector fields $Y^{(\alpha)}$ closing on a finite-dimensional real Lie 
algebra. They will be called Lie (or even Lie--Scheffers) systems. 
Many of its applications in physics and mathematics have been developed 
by Winternitz and coworkers \cite{And80}--\cite{HavPosWin99}.

\section{Some examples}

We have mentioned in the Introduction two types of systems 
of differential equations whose general solution can be written 
as described by Theorem~\ref{Lie_Theorem}: homogeneous linear systems like
 \begin{equation}
\frac {dx^i}{dt}=\sum_{j=1}^nA^i\,_j(t)\, x^j\ , \quad i=1,\ldots,n\,,
\label{hls}
\end{equation}
for which $m=n$ and the (linear) superposition function is given by 
(\ref{lsr}), and the inhomogeneous ones,
\begin{equation}
\frac {dx^i}{dt}=\sum_{j=1}^nA^i\,_j(t)\, x^j+B^i(t)\ , \quad i=1,\ldots,n \ ,\label{ils}
 \end{equation}
for which $m=n+1$ and  the (affine) superposition function is
(\ref{asr}).

In the first case, the linear  system
can be considered as the one giving the
integral curves of the $t$-dependent vector field
 \begin{equation}
 X= \sum_{i,j=1}^nA^i\,_j(t)\, x^j\,\pd{}{x^i}\ , \label{vfhs}
\end{equation}
which is a
linear combination with time-dependent coefficients, 
 \begin{equation}
X= \sum_{i,j=1}^nA^i\,_j(t)\,X_{ij}\ , \label{livfis}
 \end{equation}
of the $n^2$ vector fields
 \begin{equation}
 X_{ij}=x^j\,\pd{}{x^i}\ , \qquad i,j=1,\ldots,n\,.\label{xij}
\end{equation}
Notice that
$$ 
 [X_{ij},X_{kl}]=\left[x^j\pd {}{x^i},x^l\pd {}{x^k}\right]
=\delta^{il}\,x^j
\pd {}{x^k}-\delta^{kj}\,x^l\pd {}{x^i}\ ,
$$
i.e.  
\begin{equation}
[X_{ij},X_{kl}]=\delta^{il}\,X_{kj}-\delta^{kj}\,X_{il}\ ,
\end{equation}
which means that the vector fields $\{X_{ij}\}$, with $ i,j=1,\ldots,n$,
appearing in the case of a homogeneous system, 
close on a $n^2$-dimensional Lie algebra isomorphic 
to the ${\goth{gl}}(n,{\R})$ algebra.
It suffices to compare these commutation relations with those  of
the ${\goth{gl}}(n,{\R})$ algebra. The latter is generated by the
matrices $E_{ij}$ with elements 
$(E_{ij})_{kl}=\delta _{ik} \, \delta_{jl}$, which satisfy
$$ 
[E_{ij},E_{kl}]=\delta_{jk}\,E_{il}-\delta_{il}\,E_{kj}\ .  
$$
Therefore, in this homogeneous linear case $r=n^2$ and $m=n$, hence the 
inequality $r\leq m\,n$ is actually an equality.
 
For the case of the inhomogeneous system (\ref{ils}), the time-dependent vector field is
 \begin{equation}
X= \sum_{i=1}^n\left( \sum_{j=1}^nA^i\ _j(t)\, x^j+B^i(t)\right)\pd{}{x^i}\ , \label{vfinhs}
\end{equation}
which is a linear combination with $t$-dependent coefficients, 
 \begin{equation}
X= \sum_{i,j=1}^nA^i\,_j(t)\,X_{ij}+\sum_{i=1}^n B^i(t)\, X_i\ , \label{avfis}
\end{equation}
of the $n^2$ vector fields (\ref{xij}) and the $n$ vector fields
 \begin{equation} 
X_i=\pd{}{x^i}\ , \qquad i=1,\ldots,n\,. 
\end{equation}
Now, these last vector fields commute among themselves
$$
[X_{i},X_k]=0\ ,\qquad \forall\, i,k=1,\ldots,n\,, $$
and
$$
[X_{ij},X_k]=-\delta_{kj}\, X_i\ ,\qquad \forall\, i,j,k=1,\ldots,n\,.
$$
Therefore, the Lie  algebra generated by the vector 
fields $\{X_{ij}, X_k\mid i,j,k=1,\ldots,n\}$ is
isomorphic to the $(n^2+n)$-dimensional Lie algebra of the affine
group. In this case $r=n^2+n$ and $m=n+1$ and the 
equality $r=m\, n$ also follows.

Another remarkable example, with many applications in physics, 
is that of the Riccati equation, which corresponds to $n=1$
\cite{Win83,CarMarNas,CarRam}:
 \begin{equation}
\frac{dx(t)}{dt}=c_2(t)\,x^2(t)+c_1(t)\,x(t)+ c_0(t)\ .
\label{Riceq} 
\end{equation}
In this case $r=3$ and
$$
\xi^1(x)=1\,,\quad\xi^2(x)=x\,,\quad\xi^3(x)=x^2\,,
$$ 
while
$$
Z_1(t)=c_0(t)\,,\quad Z_2(t)=c_1(t)\,,\quad Z_3(t)=c_2(t)\ . 
$$
The equation (\ref{Riceq}) determines the integral curves of 
the $t$-dependent vector field
$$
X(x,t)=c_2(t)\,Y^{(3)}+c_1(t)\,Y^{(2)}+c_0(t)\,Y^{(1)}\ ,
$$
where the vector fields $Y^{(1)}$, $Y^{(2)}$, and $Y^{(3)}$ in  
the decomposition
are given by 
\begin{equation}
Y^{(1)} =\pd{}{x}\,,        \quad
Y^{(2)} =x\,\pd{}{x}\, ,    \quad
Y^{(3)} = x^2\,\pd{}{x}\,.  \label{Ricvf}
\end{equation}
It is quite easy to check that they close
on the following  three-dimensional real Lie  algebra,
\begin{equation}
[Y^{(1)},Y^{(2)}] = Y^{(1)}\,,      \quad
[Y^{(1)},Y^{(3)}] = 2\,Y^{(2)}\,,     \quad
[Y^{(2)},Y^{(3)}] = Y^{(3)} \,,         \label{comm_sl2}
\end{equation}
i.e. the ${\goth{sl}}(2,{\R})$ algebra.

It can be shown that, for the Riccati equation, 
$m=3$, and hence, as $r=3$  the equality $r=m\,n$ holds.
The superposition function comes
from the relation
 \begin{equation}
\frac{x-x_1}{x-x_2}:\frac{x_3-x_1}{x_3-x_2}=k\ ,
\label{superp_formula}
\end{equation}
or, in other words \cite{CarMarNas},
 \begin{equation}
x=\frac {x_1(x_3-x_2)+k\,x_2(x_1-x_3)}{(x_3-x_2)+k\,(x_1-x_3)}\ .\label{sfRe}
\end{equation} 
In particular, the solutions $x_1$, $x_2$ and $x_3$ are obtained
for $k=0$, $\infty$,  and $1$, respectively.

We show next a last example of nonlinear superposition formula 
for a specific Lie system, in order to illustrate how complicated the explicit 
expressions can become. For the sake of brevity, we give only the result. 

Consider the differential equation system 
\begin{eqnarray}
&&\frac{dx(t)}{dt}=b_1(t)+b_2(t)\,x+b_3(t)\,(x^2-y^2)\,, \nonumber\\    
&&\frac{dy(t)}{dt}=b_2(t)\,y+2 b_3(t)\,x y\,,                   \label{non_lin2}
\end{eqnarray}
which determines the integral curves of the $t$-dependent vector field
$$
X(x,t)=b_1(t)\,Y^{(1)}+b_2(t)\,Y^{(2)}+b_3(t)\,Y^{(3)}\ ,
$$
where now $Y^{(1)}$, $Y^{(2)}$, and $Y^{(3)}$ are given by   
\begin{equation}
Y^{(1)} =\pd{}{x}\,,        \quad
Y^{(2)} =x\,\pd{}{x}+y\,\pd{}{y}\, ,    \quad
Y^{(3)} = (x^2-y^2)\,\pd{}{x}+2 xy\,\pd{}{y}\,.  \label{Ric_comvf}
\end{equation}
These vector fields satisfy the commutation relations (\ref{comm_sl2}), 
hence the previous system is a Lie system with associated Lie algebra 
${\goth{sl}}(2,{\R})$, and with $r=3$, $n=2$. The number of particular 
solutions needed is $m=3$. In fact, suppose we know a set of particular
solutions $x_{(i)}=\{x_i,\,y_i\}$, $i=1,\,2,\,3$ of the system 
(\ref{non_lin2}). Then the general solution can be written as
\begin{eqnarray}
&&x=F_1(x_{(1)},\,x_{(2)},\,x_{(3)},\,k_1,\,k_2)=\frac{N_x}{D}\,,     \nonumber\\
&&y=F_2(x_{(1)},\,x_{(2)},\,x_{(3)},\,k_1,\,k_2)=\frac{N_y}{D}\,,     \label{sup_form_big}
\end{eqnarray}
where 
\begin{eqnarray}
&&N_x=x_1\{(x_2-x_3)^2+(y_2-y_3)^2\}                                    \nonumber\\
&&\quad\quad+k_1 \{x_2^2\,x_3+(x_3-x_2)x_1^2+(y_1-y_2)^2 x_3            \nonumber\\
&&\quad\quad\quad\quad-x_2(x_3^2+(y_1-y_3)^2)
-x_1((x_2-x_3)^2+(y_2-y_3)^2)\}                                         \nonumber\\
&&\quad\quad+k_2 \{x_3^2(y_2-y_1)+x_2^2\,(y_1-y_3)
+(y_3-y_2)(x_1^2+(y_1-y_2)(y_1-y_3))\}                                  \nonumber\\
&&\quad\quad+(k_1^2+k_2^2) x_2\{(x_1-x_3)^2+(y_1-y_3)^2\}\,,            \nonumber\\
&&\quad\                                                                \nonumber\\
&&N_y=y_1\{(x_2-x_3)^2+(y_2-y_3)^2\}                                    \nonumber\\
&&\quad\quad+k_1 \{x_2^2 (y_3-y_1)-x_3^2 (y_1+y_2)
+2 x_2 (x_3 y_1-x_1 y_3)                                                \nonumber\\
&&\quad\quad\quad\quad+2 x_1 x_3 y_2-(x_1^2+(y_1+y_2)(y_1-y_3))(y_2-y_3)\} \nonumber\\
&&\quad\quad+k_2\{x_1^2(x_2-x_3)+x_2^2\,x_3+x_3(y_2^2-y_1^2)            \nonumber\\
&&\quad\quad\quad\quad-x_2(x_3^2+y_3^2-y_1^2)+x_1(x_3^2-x_2^2+y_3^2-y_2^2)\} \nonumber\\
&&\quad\quad+(k_1^2+k_2^2) y_2 \{(x_1-x_3)^2+(y_1-y_3)^2\}\,,           \nonumber\\
&&\quad\                                                                \nonumber\\
&&D=(x_2-x_3)^2+(y_2-y_3)^2                                           \nonumber\\
&&\quad\quad-2 k_1\{ (x_1-x_3)(x_2-x_3)+(y_1-y_3)(y_2-y_3) \}           \nonumber\\
&&\quad\quad+2 k_2\{ x_3(y_2-y_1)+x_2(y_1-y_3)+x_1(y_3-y_2)\}           \nonumber\\
&&\quad\quad+(k_1^2+k_2^2)\{(x_1-x_3)^2+(y_1-y_3)^2\}\,,                \nonumber
\end{eqnarray}
and $k_1$, $k_2$ are two arbitrary real constants determining each
particular solution. For example, the particular solutions 
$\{x_1,\,y_1\}$, $\{x_2,\,y_2\}$ and $\{x_3,\,y_3\}$ can be obtained by taking
$k_1=k_2=0$, the limit $k_1\rightarrow\infty$ (or $k_2\rightarrow\infty$), 
and $k_1=1$, $k_2=0$, respectively.

In particular, if we look for solutions of the system (\ref{non_lin2}) 
with $y=0$, we recover, essentially, the Riccati equation (\ref{Riceq}); 
likewise, the superposition formula (\ref{sup_form_big}) 
reduces to (\ref{sfRe}) in such a particular case.

For a more complete information about the explicit construction and use of 
superposition formulas, see, e.g., \cite{CGM}--\cite{CarMarNas} and 
references therein.

\section{Lie--Scheffers systems on Lie groups}

The most important  example, which will be shown to give rise to many other related systems, 
occurs when $M$ is a Lie group $G$ and we consider vector fields $X_\alpha$ in $G$
that are either left-invariant or right-invariant as corresponding  
either to the Lie algebra $\goth g$ of $G$ or to the 
opposite algebra \cite{CarRamGra,CGM01}.

Let us choose a basis $\{a_1,\ldots,a_r\}$ for the tangent space
$T_eG$ at the neutral element $e\in G$, and denote 
$\{\vartheta_1,\ldots,\vartheta_r\}$ 
the corresponding dual basis of $T_e^*G$. In the following 
  $X^R_\alpha$ denotes the right-invariant vector field in $G$ such that 
$X^R_\alpha(e)=a_\alpha$, i.e.
$$X^R_\alpha (g)=R_{g*e}(a_\alpha) \ ,$$ 
and in an analogous way, $X^L_\alpha$ will denote the left-invariant vector
 field  
$$X^L_\alpha (g)=L_{g*e}(a_\alpha)\ .$$
Similarly, $\theta^R_\alpha$ and $\theta^L_\alpha$ are the right- and left-invariant 1-forms in $G$ determined by $\vartheta_\alpha$, i.e.
$$ \theta^R_\alpha(g)=(R_{g^{-1}})^*_e(\vartheta_\alpha)\ ,\qquad 
\theta^L_\alpha(g)=(L_{g^{-1}})^*_e(\vartheta_\alpha)\ .
$$

If we consider the right-invariant Lie--Scheffers system on $G$
\begin{equation}
X(g,t)=-\sum_{\alpha=1}^r b_\alpha(t)\,X^R_\alpha(g)\ , \label{Xrightinv}
\end{equation}
its integral curves will be determined by the system of 
differential equations 
\begin{equation}
\dot g(t)=-\sum_{\alpha=1}^rb_\alpha(t)X^R_\alpha(g(t))\ .\label{icXrightinv}
\end{equation}
Applying $(R_{g(t)^{-1}})_{*g(t)}$ to both sides of this
equation  we  obtain the equivalent equation
\begin{equation}
(R_{g(t)^{-1}})_{*g(t)}(\dot
g(t))=-\sum_{\alpha=1}^rb_\alpha(t)\,a_\alpha\ ,  \label{LSingr}
\end{equation}
which we will write as well, with a slight abuse of notation, as
\begin{equation}
(\dot g\, g^{-1})(t) =-\sum_{\alpha=1}^rb_\alpha(t)a_\alpha\ ,\label{eqingr}
\end{equation}
although (\ref{LSingr}) reduces to (\ref{eqingr}) only when $G$ is a 
matrix group.
This equation is right-invariant, and so, out of a solution $\bar g(t)$  of
 (\ref{LSingr})  with initial condition
 $\bar g(0)=e$, the solution with initial conditions $g(0)=g_0$ is given by 
$\bar g(t)g_0$. This means that for the  Lie--Scheffers system (\ref{Xrightinv})
on the Lie group $G$, $m=1$.

Of course, given a homomorphism of Lie groups $F:G\to G'$, the right-invariant
Lie--Scheffers system on $G$  (\ref{Xrightinv}) produces a  right-invariant
Lie--Scheffers system on $G'$,
$$
X(g',t)=-\sum_{\alpha=1}^r b_\alpha(t)\,(F_*X)^R_\alpha(g')\ ,
$$
where $(F_*X)^R_\alpha$ is the right-invariant vector field on $G'$ which is 
$F$-related with the vector field  $X^R_\alpha$.

Let us  consider a left action of a Lie
group $G$, with Lie algebra $\goth g$, on a manifold $M$, 
$\Phi:G\times M\to M$, and denote $\Phi_g:M\to G$ and $\Phi_p:G\to G$, 
where $g\in G$, $p\in M$,  the maps defined by 
$\Phi_g(p)=\Phi_p(g)=\Phi(g,p)$.  The fundamental vector field
$X_a$  associated to the element $a$ of~${\goth g}$ is
given  by
$$
(X_a f)(p) = \frac d{dt} f(\Phi(\exp(-ta),p)) \Bigr|_{t=0} \,,
\quad f\in C^\infty(M)\ ,
$$
where  the minus sign has been introduced for $X:{\goth g}\to \X(M)$ 
to be a Lie algebra homomorphism, i.e. a $\R$-linear map such 
that $X_{[a,b]} = [X_a,X_b]$. Note that
$X_a(p)=\Phi_{p*e}(-a)$.
These vector fields are always complete. 
As an example, let us consider 
the left action  action of $G$ on itself by left translations, 
$\Phi(g,g')=g\,g'$. The fundamental vector fields $X_a$ are right invariant
because 
$$ 
(X_a)(g)=\Phi_{g*e}(-a)=R_{g*e}(-a)=-(X^R_a)(g)\ ,
$$ 
where $X^R_a$ is the right-invariant vector field in $G$ 
determined by its value at the neutral element  $(X^R_a)(e)=a$.
 
Given two actions  $\Phi_1$ and  $\Phi_2$ of a Lie group $G$ on two
differentiable  manifolds
$M_1$ and  $M_2$, a map $F:M_1\to M_2$ is said to be 
equivariant 
 if  $F\circ \Phi_{1g}=\Phi_{2g}\circ F$. The remarkable property is that
when $G$ is connected, the map 
 $F:M_1\to M_2$ is equivariant 
if and only if for each  $a\in T_eG$ the
corresponding fundamental vector fields in $M_1$ and $M_2$ are 
$F$-related \cite{CGM,CarRamGra,CGM01}.

Now, let $H$ be a closed subgroup of $G$ and consider the homogeneous space $M=G/H$.
Then, $G$ acts on $M$ by $\lambda(g',gH)=(g'g)H$. Moreover, $G$  
can be seen as a principal bundle $(G,\tau,G/H)$ over $G/H$, 
where $\tau$ denotes the canonical projection. 
The important point is (see e.g. \cite{CarRamGra}) 
that the map $\tau:G\to G/H$ is equivariant, with respect to the 
left action of $G$ on itself by left translations 
and the action $\lambda$ on $G/H$,  
and consequently, the fundamental vector fields 
corresponding to the two actions 
are $\tau$-related. Therefore, 
the right-invariant vector fields $X^R_\alpha$
are $\tau$-projectable and 
the $\tau$-related vector fields in $M$ are the fundamental vector fields 
$-X_\alpha=-
X_{a_\alpha}$
corresponding to the natural left action of $G$ on $M$,
$\tau_{*g}X_\alpha^R(g)=-X_\alpha(gH)$. In this way we 
will have an associated Lie--Scheffers system on $M$:
\begin{equation}
X(x,t)=\sum_{\alpha=1}^rb_\alpha(t)X_\alpha(x)\ ,\label{LSsystem}
\end{equation}
where $x=gH$, whose integral curve will be determined by 
\begin{equation}
\dot x=\sum_{\alpha=1}^rb_\alpha(t)X_\alpha(x)\ .\label{eqLSsystem}
\end{equation}
Thus, the solution of (\ref{LSsystem}) starting from $x_0$ will be
$x(t)=\Phi(g(t),x_0)$, with $g(t)$ being the  solution of 
(\ref{LSingr}) with $g(0)=e$. This is the main point:
the knowledge of one particular solution of (\ref{LSingr}) allows us
to obtain the general solution of (\ref{LSsystem}).

The converse property is true in the following sense:
Given a Lie--Scheffers system in a manifold $M$ 
defined by complete vector fields
 and with associated Lie algebra $\goth g$, we can see these
as fundamental vector fields relative to an action given by integrating 
the vector fields. Then, the restriction to an orbit will provide a
 homogeneous space of the above type. The choice of a point $x_0$ in the 
homogeneous space allows us to identify the homogeneous space $M$ with $G/H$,
 where $H$ is the stability group of $x_0$. Different choices for $x_0$ will 
lead to conjugate subgroups \cite{CarRamGra}. 

For instance, the vector fields appearing in (\ref{Ricvf}) close on a Lie 
algebra but the third one is not complete on $\R$. We can however consider 
the one-point compactification of $\R$, 
${\overline {\R}}={\R}\cup\{\infty\}$, 
and then the flows of vector fields in  (\ref{Ricvf}) are, respectively,
$$
x\mapsto x+\epsilon\,,\quad x\mapsto e^\epsilon x\,,\quad x
\mapsto\frac x{1-x\,\epsilon}\ ,
$$
and therefore they can be considered as the fundamental vector fields 
corresponding to the action of $SL(2,{\R})$ on the completed real
line $\overline{\R}$, given by \cite{CarRam} 
\begin{eqnarray}
\Phi(A,x)={\frac{\alpha\, x+\beta}{\gamma\, x+\delta}},\ \ \ \mbox{if}\
x\neq-{\frac{\delta}{\gamma}},
\nonumber\\
\Phi(A,\infty)={\frac{\alpha}{\gamma}}\ ,\ \ \ \
\Phi\left(A,-{\delta}/{\gamma}\right)=\infty,
\nonumber\\
\mbox{when}\ \ \ A=\matriz{cc}{\alpha&\beta\\\gamma&\delta}\,\in SL(2,{\R}).
\label{action_SL2}
\end{eqnarray}

The stability group of $\infty$ is the subgroup of matrices with $\gamma=0$,
 which is isomorphic to the affine group ${\mathcal{A}}_1$ 
in one dimensional space, while the stability group of $0$ is made up by the
 matrices with $\beta=0$, a group isomorphic to ${\mathcal{A}}_1$. Indeed 
$$
\matriz{cc}{\delta&0\\-\gamma&\alpha}
=\matriz{cc}{0&1\\-1&0}
\matriz{cc}{\alpha&\gamma\\0&\delta}
\matriz{cc}{0&-1\\1&0}\ .
$$

The remarkable fact is that equation (\ref{LSingr}) has a universal character. 
There will be many Lie--Scheffers
 systems 
associated with such an equation. It is enough to consider homogeneous spaces
and the corresponding fundamental vector fields. In this way we will get a 
set of different systems corresponding to the same equation on the Lie group $G$. 
In particular, we can consider an action of $G$ on a linear space given by a
linear representation, and then the associated Lie systems are linear systems. 
Hence, we obtain a kind of linearization of the original problem \cite{Win83}. 
Therefore, the theory can  be useful in the study of both 
classical and quantum problems.

As an example we can consider both the Riccati equation
$$
\dot x=b_0(t)+2\, b_1(t)\,x+b_2(t)\,x^2\ ,
$$
and the linear system of first order differential equations
$$
\frac d{dt}\matriz{c}{x\\y}=\matriz{cc}{b_1(t)&b_0(t)\\-b_2(t)&-b_1(t)}
\matriz{c}{x\\y}\ .
$$
They are two different Lie systems for $G=SL(2,\R)$ corresponding 
to the same equation 
$$
\dot g g^{-1}=-b_0(t)\,M_0-2\,b_1(t)\,M_1-b_2(t)\,M_2\ ,
$$
with the matrices
$$
M_0=\matriz{cc}{0&1\\0&0}\,,
\quad M_1=\frac{1}{2}\matriz{cc}{1&0\\0&-1}\,,\quad M_2=\matriz{cc}{0&0\\-1&0}\,,
$$
being a basis of ${\goth{sl}}(2,\R)$. 
They  satisfy the commutation relations
$$
[M_0,M_1]=-M_0\,,\quad [M_0,M_2]=-2\, M_1\,,\quad [M_1,M_2]=-M_2\ .
$$

As another illustrative example, we can consider the 
non-relativistic dynamics of  a spin $1/2$ particle, when only the
spinorial part is considered \cite{CGM01}. 
The dynamics
of such a particle in a time-dependent magnetic field is
described by the so called Schr\"odinger--Pauli equation:
$$
i\hbar\frac{d\ket\psi}{dt}=H\ket\psi =-\mu\,\vec B\cdot \vec
S\ket\psi\ ,\label{SPeq}
$$
with $\mu$  proportional to the Bohr magneton, $\vec B=(B^1, B^2,B^3)$  the
$t$-dependent magnetic field, and  $S_i=\frac\hbar 2\,\sigma_i$.
More explicitly, 
\begin{equation}
\frac d{dt} \matriz{c}{\psi_1\\\psi_2}
= \frac \mu 2\matriz{cc}{i B^3&iB^1+B^2\\ iB^1-B^2& -iB^3}
\matriz{c}{\psi_1\\\psi_2}\ .                           \label{SPeqdos}
\end{equation}

The matrices $-i\sigma^1$,  $-i\sigma^2$ and $-i\sigma^3$ generate the
real Lie algebra of traceless skew-Hermitian $2\times 2$ matrices,
the Lie algebra of the group $SU(2,\C)$ and therefore of $SO(3,\R)$. 

As a consequence of the theory we have developed in this section, 
in order to find the general solution of the evolution equation (\ref{SPeqdos}), 
it suffices to determine  the curve
$R(t)$ in $SO(3,\R)$ starting from the identity map, $R(0)=I$  and  such that
$$\dot R\,R^{-1}=B^3M_3+B^2M_2+B^1M_1 \ ,  $$
where
$$
M_1=\matriz{ccc}{0&0&0\\0&0&-1\\0&1&0}\,,
\quad M_2=\matriz{ccc}{0&0&1\\0&0&0\\-1&0&0}\,,
\quad M_3=\matriz{ccc}{0&-1&0\\1&0&0\\0&0&0}\,.
$$
Such a curve gives us
the general solution for the dynamics
$$
\ket{\psi(t)}=\bar R(t)\ket{\psi(0)}\ ,  
$$
where $\bar R$ is an element in $SU(2,\C)$ corresponding to $R$.

\section{The Wei--Norman method\label{Wei_Nor_meth}}

In order  to solve directly the equation (\ref{LSingr}) we can use a method 
which is 
a generalization of the one proposed by 
Wei and Norman \cite{WN1,WN2} for finding the time evolution operator 
for a linear systems of type
${dU(t)}/{dt}=H(t)U(t)$, with $U(0)=I$, see also \cite{CarMarNas}. 
However, as it will be mentioned in a later section, there exist 
alternative methods for solving (\ref{LSingr}) by reducing
 the problem to a simpler one.
 
Both procedures are based on the following property \cite{CarRamGra}: 
If $g(t)$, $g_1(t)$ and $g_2(t)$ are differentiable curves in
$G$ such that  $g(t)=g_1(t)g_2(t)$, $\forall t\in{\mathbb R}$, then, 
\begin{equation}
R_{g(t)^{-1}\,*g(t)}(\dot g(t))
=R_{g_1(t)^{-1}\,*g_1(t)}(\dot g_1(t))                   
+\Ad(g_1(t))\left\{R_{g_2(t)^{-1}\,*g_2(t)}(\dot g_2(t))\right\}\,. 
\label{1cocycle_fcanr}
\end{equation}

The generalization of this property to several factors is as follows. 
Let now $g(t)$ be a  curve in $G$ which is given by the product of other
$l$ curves $g(t)=g_1(t)g_2(t)\cdots g_l(t)=\prod_{i=1}^l g_i(t)$.
Then, denoting $h_s(t)=\prod_{i=s+1}^{l} g_i(t)$, for $s\in\{1,\,\dots,\,l-1\}$,
and applying (\ref{1cocycle_fcanr}) to $g(t)=g_1(t)\, h_1(t)$ we have
\begin{eqnarray}
R_{g(t)^{-1}\,*g(t)}(\dot g(t))
&=&R_{g_1(t)^{-1}\,*g_1(t)}(\dot g_1(t))   
+\Ad(g_1(t))\left\{R_{h_1(t)^{-1}\,*h_1(t)}(\dot h_1(t))\right\}\,. \nonumber
\end{eqnarray}
Simply iterating, and using that $\Ad(g g^\prime)=\Ad(g)\Ad(g^\prime)$ for all
$g,\,g^\prime\in G$ we obtain
\begin{eqnarray}
R_{g(t)^{-1}\,*g(t)}(\dot g(t))
&=&R_{g_1(t)^{-1}\,*g_1(t)}(\dot g_1(t))                        
+\Ad(g_1(t))\left\{R_{g_2(t)^{-1}\,*g_2(t)}(\dot g_2(t))\right\} \nonumber\\
&&\quad+\cdots+\Ad\left(\prod_{i=1}^{l-1} g_i(t)\right)
\left\{R_{g_l(t)^{-1}\,*g_l(t)}(\dot g_l(t))\right\}            \nonumber\\
&=&\sum_{i=1}^l \Ad\left(\prod_{j<i} g_j(t)\right)
\left\{R_{g_i(t)^{-1}\,*g_i(t)}(\dot g_i(t))\right\}            \nonumber\\
&=&\sum_{i=1}^l \left(\prod_{j<i} \Ad(g_j(t))\right)
\left\{R_{g_i(t)^{-1}\,*g_i(t)}(\dot g_i(t))\right\}\,,         
\label{iter_coc}
\end{eqnarray}
where it has been taken $g_0(t)=e$ for all $t$.

The generalized  Wei--Norman method consists on writing
the solution $g(t)$ of (\ref{LSingr}) 
in terms of its second kind canonical coordinates 
w.r.t. a basis $\{a_1,\,\dots,\,a_r\}$ of the Lie 
algebra $\goth g$, for each value of  $t$, i.e.
$$
g(t)=\prod_{\alpha=1}^{r}\exp(-v_\alpha(t)a_\alpha)=\exp(-v_1(t)a_1)\cdots
 \exp(-v_r(t)a_r)\ ,
$$
and transforming the differential equation (\ref{LSingr})
into a differential equation system for the $v_\alpha(t)$,
with initial conditions $v_\alpha(0)=0$ for all 
$\alpha=1,\,\dots,\,r$. 
The minus signs in the exponentials have been introduced for 
computational convenience. 
Then, we use the result (\ref{iter_coc}),
taking $l=r=\mbox{dim}\,G$ and  
$g_\alpha(t)=\exp(-v_\alpha(t) a_\alpha)$ for all $\alpha$. 
Now, since 
$R_{g_\alpha(t)^{-1}\,*g_\alpha(t)}(\dot g_\alpha(t))
=-\dot v_\alpha(t) a_\alpha$, we see that 
(\ref{iter_coc}) reduces to  
\begin{eqnarray}
R_{g(t)^{-1}\,*g(t)}(\dot g(t))
&=&-\sum_{\alpha=1}^r \dot v_\alpha \left(\prod_{\beta<\alpha} 
\Ad(\exp(-v_\beta(t) a_\beta))\right)a_\alpha           \nonumber\\
&=&-\sum_{\alpha=1}^r \dot v_\alpha \left(\prod_{\beta<\alpha} 
\exp(-v_\beta(t) \ad(a_\beta))\right)a_\alpha\,,                \nonumber
\end{eqnarray}
where it has been used the identity $\Ad(\exp(a))=\exp(\ad(a))$, for all
$a\in\goth g$.
Substituting in equation (\ref{LSingr}) we 
obtain the fundamental expression of the Wei--Norman method 
\begin{equation}
\sum_{\alpha=1}^r \dot v_\alpha \left(\prod_{\beta<\alpha} 
\exp(-v_\beta(t) \ad(a_\beta))\right)a_\alpha
=\sum_{\alpha=1}^r b_\alpha(t) a_\alpha\,,
\label{eq_met_WN}
\end{equation}
with  $v_\alpha(0)=0$, $\alpha=1,\,\dots,\,r$.
The resulting differential equation
system for the functions $v_\alpha(t)$ is integrable 
by quadratures if the Lie algebra is solvable 
\cite{WN1,WN2}, and in particular, for nilpotent 
Lie algebras. 

As a simple but illustrative example 
we can consider the affine group
in one dimension,  ${\cal A}_1$,  i.e. the set of 
transformations of the real line 
\begin{equation}
\bar x=\alpha_1\, x+\alpha_0\ ,\label{acta1}
\end{equation} 
with $\alpha_1\ne 0$ and $\alpha_0$ being real
numbers. The group composition law is 
$$
(\alpha'_0,\alpha'_1)*(\alpha_0,\alpha_1)
=(\alpha'_0+\alpha'_1\,\alpha_0,\alpha'_1\,\alpha_1)\ .
$$ 
Denoting by $(x_0,x_1)$ the coordinate system  in   ${\cal A}_1$ given by
$$
x_0(\alpha_0,\alpha_1)=\alpha_0\,,\qquad  
x_1(\alpha_0,\alpha_1)=\alpha_1\ ,
$$ 
we see that a basis of
right-invariant vector fields in  ${\cal A}_1$ is given 
by 
$$
X^R_0=\pd{}{x_0}\,,\qquad X_1^R=x_0\,\pd{}{x_0}+x_1\,\pd{}{x_1}\ ,
$$ 
while the corresponding basis of left-invariant vector fields 
is given by 
$$
X^L_0=x_1\,\pd{}{x_0}\,, \qquad X^L_1=x_1\,\pd{}{x_1}\ ,
$$ 
therefore the defining relations are 
$$
[a_0,a_1]=-a_0\ .
$$ 
Then, 
$$
\ad (a_0)a_0=0\,, \qquad \ad (a_0)a_1=-a_0\ ,
$$  
and
if $g=\exp(-u_0a_0)\exp(-u_1a_1)$, equation (\ref{eq_met_WN}) becomes  
in this case
$$
\dot u_0\,a_0+\dot u_1(a_1+u_0\,a_0)=b_0\, a_0+b_1\, a_1\,,     
$$
so we obtain the system
\begin{equation}
\dot u_0=b_0-b_1\,u_0\,, \qquad  \dot u_1=b_1\ ,
\label{sis_red_1}
\end{equation}
with the initial conditions $u_0(0)=u_1(0)=0$. 
Note that the first equation is 
nothing but an inhomogeneous linear equation.
The explicit solution can be obtained through two
quadratures:  
$$
u_0(t)=e^{-\int_0^tdt'\, b_1(t')} \int_0^tdt'\, b_0(t')\, e^{\int_0^{t'}dt''\, b_1(t'')}\,,\qquad 
u_1(t)=\int_0^tdt'\, b_1(t')\,,
$$

In a similar way, if we consider instead
$g=\exp(-v_1a_1)\exp(-v_0a_0)$, and we take into account that 
$$\ad (a_1)a_0=a_0\,, \qquad \ad (a_1)a_1=0\ ,$$
then we will find 
$$
\dot v_1\,a_1 +\dot v_0\exp(-v_1\ad(a_1))a_0
=\dot v_1 \,a_1 +\dot v_0 \,e^{-v_1}\,a_0
=b_0\, a_0+b_1\, a_1\,,     
$$
yielding the system
\begin{equation}
\dot v_0= b_0\, e^{v_1}\,,\qquad \dot v_1=b_1\ ,
\label{sis_red_2}
\end{equation}
also with the initial conditions $v_0(0)=v_1(0)=0$. The system  
(\ref{sis_red_2}), with such initial conditions,
can be easily integrated by
two quadratures:
$$
v_0(t)=\int_0^tdt'\, b_0(t')\, e^{\int_0^{t'}dt''\, b_1(t'')}\,,\qquad 
v_1(t)=\int_0^tdt'\, b_1(t')\ .
$$

When we consider the previous action (\ref{acta1}) on the real line 
we get as fundamental vector fields 
$X_0=-\pd{}x$ and $X_1=-x\pd{}{x}$, thus the Lie system in $\mathbb R$ which
corresponds to $\dot g \, g^{-1}=-b_0\, a_0-b_1\, a_1$ 
is $\dot x=-b_0-b_1\, x$. Our theory gives us the formula for the 
explicit general solution of such inhomogeneous linear
differential equation, by making use of 
\begin{eqnarray}
x(t)&=&\Phi(g(t),x_0)=\Phi(\exp(-u_0(t)a_0),\Phi(\exp(-u_1(t)a_1),x_0)) \nonumber\\
&=&e^{-u_1(t)} x_0-u_0(t)\ ,                                            \nonumber
\end{eqnarray}
namely
\begin{equation}
x(t)=e^{-\int_0^{t}dt'\, b_1(t')}
\left\{x_0-\int_0^{t} dt'\,b_0(t')\,e^{\int_0^{t'}dt''\, b_1(t'')}\right\}\ .
\label{sol_g_x}
\end{equation}
Likewise, the same solution can be obtained from the second 
factorization:  
\begin{eqnarray}
x(t)=\Phi(\exp(-v_1(t)a_1),\Phi(\exp(-v_0(t)a_0),x_0))=e^{-v_1(t)}(x_0-v_0(t))\ ,
\end{eqnarray}
which clearly gives the same result.

\section{Connections and Lie systems}

If $G$ is a connected Lie group, the set of curves 
\begin{eqnarray}
\gamma&:&{\mathbb R}\to G\cr
&&t\mapsto g(t)\nonumber
\end{eqnarray}
is also a  group when the following composition law is considered:
$$\gamma_2*\gamma_1:t\mapsto g_2(t)\,g_1(t)\ .
$$ 

Given a curve $\gamma$ in $G$,  $g(t)$, such that 
$g(0)=e$, then 
$\bar g(t)=g(t)\,g_0$  is another  curve in $G$ starting from 
 $g_0$ (it is said to be right translated of $\gamma$ by $g_0$) and similarly
$\bar {\bar g}(t)=g_0\,g(t)$  is also a curve in $G$  starting from $g_0$ 
(called left-translated from $\gamma$ by 
$g_0$).

Now, given the curve $\gamma$, we have a vector field along $\gamma$ given
 by the tangent vector $\dot g(t)$, and then, translating these tangent vectors
to the neutral element by $R_{g^{-1}(t)*g(t)}$ we obtain a curve in 
$T_eG$ like in (\ref{eqingr}). The curve $\bar g(t)=g(t)\,g_0$, 
right-translated by $g_0$ of $\gamma$, gives rise to the same equation.

But we can consider such  equation (\ref{eqingr}) as an equation for the curve $g(t)$ 
in $G$ determined by the curve
 $a(t)=-\sum_{\alpha=1}^r b_\alpha(t)\,a_\alpha$ 
in $T_eG$. This equation is right-invariant in the sense that if 
 $g(t)$ is a solution such that 
 $g(0)=e$, then, for each  $g_0\in G$, 
$\bar g(t)=g(t)\,g_0$  is a new solution, now such that 
$\bar g(0)=g_0$.

We remark that  if $G$ is a Lie group,  
$\pi_2:P=G\times {\Bbb R}\to {\Bbb R}$ 
defines a principal $G$-bundle. The right action of $G$  on $P$ 
is given by  
$$\Psi((g',t),g)=\Psi_g((g',t))=(g'\,g,t)\ ,$$
i.e. $\Psi_g=R_g\times \id_{\mathbb{R}}$.

Giving a connection in $P$ is equivalent to give a curve 
in  $G$, for instance, one such that 
 $g(0)=e$.  It is also well-known that each global section 
provides a different trivialization of the principal bundle $P$. 
The given curve furnishes a section for  $\pi_2$, 
$\sigma(t)=(g(t),t)$, and a family of sections 
right-translated from such a  section, 
$$\{\sigma'(t)=\Psi(\sigma (t),g_0)\mid g_0\in G\}\ .$$

The tangent vectors to such family of sections span the horizontal spaces in each point.
More specifically,  horizontal and  vertical spaces in a point of 
$P$ are given by:
$$VP_{(g_0,t)}=\<(X_\alpha^R(g_0),0)>\ ,
$$
$$HP_{(g_0,t)}=\<(R_{g_0*e}(\dot g(t)\,g^{-1}(t)),1)>\ .
$$
Note that
$$
\Psi_{g^{-1}(t)*g(t)}(\dot g(t),1)
=(R_{g^{-1}(t)*g(t)}\dot g(t),1)=(\dot g(t)\,g^{-1}(t),1)\ .
$$

The choice of the connection given by $\gamma$ amounts to choose a basis 
of the tangent space at the point $(g_0,t)$ as follows,  
$\{X^R_\alpha(g_0),\partial/\partial t+R_{g_0*e}(\dot g(t)\,g^{-1}(t))\}$, 
 while the dual basis is made up by
$\{\theta^R_\alpha(g_0)-\tau_\alpha(t)\, dt,dt\}$ where the coefficients
 $\tau_\alpha(t)$ are determined by 
$$ R_{g_0*e}(\dot g(t)\,g^{-1}(t))=
\sum_{\beta} \tau_\beta(t)\, X^R_\beta(g_0)\ ,$$
i.e. $\tau_\alpha(t)=\<\theta_\alpha^R(g_0),R_{g_0*e}(\dot g(t)\,g^{-1}(t)) >$.
 Therefore, the vertical projector
 associated to the connection is 
$$v_{(g_0,t)}=\sum_{\beta} X^R_\beta(g_0)\otimes(\theta^R_\beta(g_0)-\tau_\beta
(t)\,dt )=\id_{T_{g_0}G}-\left( R_{g_0*e}(\dot g(t)\,g^{-1}(t))\right)dt\ .
$$

It is also well-known that when  a left action  $\Phi:G\times M\to M$ of 
$G$ on $M$ is considered, there exists an associated bundle $E$ with base $\R$ 
and typical fiber $M$. The total space of such bundle is the set of orbits 
of the right action of $G$ on $P\times M$,
$$(u,x)g=(\Psi(u,g),\Phi(g^{-1},x))\ ,
$$
being the projection $\pi_E[u,x]=\pi_2(u)$, where $[u,x]$ denotes 
the equivalence class of $(u,x)\in P\times M$ and $u$ is of the form $(g^\prime,t)$. 
A connection in the 
principal bundle translates into a connection in the associated bundle 
$E$, and so the horizontal curves will then be
$[\widetilde \gamma(t),x]$, where  $\widetilde \gamma(t)$ is an horizontal curve in
$P$. More explicitly,  as the curves $\widetilde \gamma$ are of the form 
$\widetilde \gamma(t)=(g(t)g_0,t)$, we find that the horizontal curves 
in the associated bundle are 
$$[(g(t)g_0,t),x_0]=[\Psi((e,t),g(t)g_0),x_0]=[(e,t), \Phi(g(t)g_0,x_0)]\ ,
$$
and consequently, 
$$[(g(t)g_0,t),x_0]=[(e,t), \Phi(g(t),\Phi(g_0,x_0))] \ .
$$

Since the principal bundle is trivial, $E$ is equivalent to a product.
When $\Phi$ is transitive, $E=M\times{\R}$, where $[(e,t),x]$ corresponds to 
$(x,t)$ and with this identification, the horizontal curve here considered 
corresponds to the integral curve starting from the 
point  $\Phi(g_0,x_0)$ of the associated Lie system 
in $M$ with respect to the action of $G$ on $M$ given by $\Phi$.

Of course the simplest case is when a linear representation of $G$ on 
the vector space $V$ is considered, the associated bundle being 
then a vector bundle
and the corresponding Lie system being a linear system.  This means that 
a system  as in (\ref{hls})
can be seen as defining the horizontal curves corresponding to a
 connection in an associated vector bundle. The fact that linear systems, as 
Schr\"odinger equations, could be thought of as defining horizontal curves of a
connection were considered several years ago \cite{ACP} and it has been suggested
recently by looking at the transformation properties of the equation under
certain gauge changes \cite{MPL}.

\section{The reduction method}

Given an equation on a Lie group 
\begin{equation}
\dot g(t)\, g(t)^{-1}= a(t)=-\sum_{\alpha=1}^rb_\alpha(t)\,a_\alpha\in T_eG \ ,
\label{eqingr2}
\end{equation}
with $g(0)=e\in G$, it may happen that the only non-vanishing coefficients
are those corresponding to a subalgebra $\goth h$ of $\goth g$. Then the 
equation reduces 
to a simpler equation on a subgroup, involving less coordinate functions
 in the Wei--Norman method.

On the other hand, we know that such an equation (\ref{eqingr2}) can be seen as a 
connection in a principal bundle, and it is also well-known that the group of 
automorphisms of the principal bundle acts on the set of connections. The
 automorphims 
of the bundle we are considering are given by curves $g'(t)$ in the group $G$, and 
so the group of curves in $G$ defines an action on the set of connections and
 therefore on the Lie systems on the group. We can take advantage of such
 an action for transforming a given Lie system in another simpler one. 

Now, let us choose a curve $g^\prime(t)$ in the group $G$, corresponding to a given automorphism,  and define 
the curve $\overline g(t)$ by $\overline g(t)=g^\prime(t)g(t)$, 
where $g(t)$ is the previous solution of (\ref{LSingr}).  The new curve in 
$G$, $\overline g(t)$, determines a new connection and therefore a
 new Lie system. 

Indeed, from (\ref{1cocycle_fcanr}),
\begin{equation}
R_{\overline g(t)^{-1}* \overline g(t)}(\dot {\overline g}(t))
=R_{g^{\prime\,-1}(t)*g^{\prime}(t)}(\dot g^\prime(t))
-\sum_{\alpha=1}^r b_\alpha(t)\Ad(g^{\prime}(t))a_\alpha\ , 
\label{eq_reducir}
\end{equation}
which is an equation similar to (\ref{LSingr})
but with different right hand side. Therefore, the aim is
to choose the curve $g^\prime(t)$ appropriately, i.e. in such a way that 
 the new equation be simpler. 
For instance, we can choose a subgroup $H$ and 
look for a choice of $g'(t)$ such that the right hand side 
of (\ref{eq_reducir}) lies in $T_e H$, and hence $\overline g(t)\in H$ for all
$t$.

Now, suppose we consider a transitive action $\Phi:G\times M\to M$ 
of $G$ on a homogeneous space $M$, 
which can be identified with the set $G/H$ of left-cosets, 
by choosing a fixed point $x_0$: $H$ is then the stability subgroup of $x_0$. 
The horizontal curves 
starting from the point $x_0$ associated to both connections are related by 
$$\overline x(t)=\Phi(\overline g(t),x_0)=\Phi(g^\prime(t)g(t),x_0)= 
\Phi(g^\prime(t),\Phi(g(t),x_0))=\Phi(g^\prime(t),x(t))\ .$$

Therefore, the action of the group of curves in $G$ on the set of connections 
translates to the homogeneous space and gives an action on the 
corresponding set of associated Lie systems. More explicitly, if 
we consider the automorphism defined by $g'(t)$, the Lie system 
(\ref{eqLSsystem})
transforms into a new one (see \cite{CarRamGra})
\begin{equation}
\dot{\bar x}=\sum_{\alpha=1}^r\bar b_\alpha(t)X_\alpha(\bar x)\ ,
\label{eqLSsystem_hom_sp}
\end{equation}
in which 
$$
\bar b=\Ad(g'(t))b(t)+\dot g'\, g^{\prime{-1}}\ .
$$

The important result   proved in \cite{CarRamGra} is that the knowledge of a 
particular solution of the associated Lie system to (\ref{LSingr}) 
in $G/H$ allows us to reduce the problem to one in the subgroup $H$.  
For any choice of the curve $g^\prime(t)$  we can consider a
 curve $x^\prime(t)$ defined in the homogeneous 
space $G/H$ by $g^\prime(t)$ as follows: 
$x^\prime(t)=\tau(g^{\prime\,-1}(t)\overline g(t))=g^{\prime\,-1}(t)H$.
Then, if  $g^\prime(t)$
is chosen such that the curve $x^\prime(t)$ is a solution of the
 associated system, then the automorphism defined by $g^\prime(t)$  
transforms the original problem into one in the subgroup $H$:
\begin{theorem} 
Every integral curve of the time-dependent vector field on the group $G$,
given by  the right hand side of (\ref{icXrightinv}), can be written in the 
form $g(t)=g_1(t)\,h(t)$, where $g_1(t)$ is a curve projecting onto a solution 
$x_1(t)$ of an equation of type (\ref{eqLSsystem}) for the natural 
left action of $G$ on the homogeneous space $G/H$,
and $h(t)$ is a solution of a type (\ref{LSingr}) equation but for the
subgroup $H$, given explicitly by
$$
(\dot h\, h^{-1})(t) 
=-\mbox{\Ad}(g_1^{-1}(t))\left(\sum_{\alpha=1}^r b_\alpha(t)a_\alpha
+(\dot g_1\,g_1^{-1})(t)\right)\in {T_eH}\ .
$$
\label{our_theorem}
\end{theorem} 

As an example, we can consider once again the affine group in one dimension, 
${\mathcal{A}}_1$, of the preceding section.
We can choose first  the Lie subgroup $H_0=\{(a_0,1)\mid a_0\in {\mathbb R}\}$ 
and consider the corresponding 
one-dimensional  homogeneous space ${\cal A}_1/H_0$.  Its
points  can be characterized
by $y=x_1$, with $x_1\ne 0$. 
In this coordinate system for ${\cal A}_1/H_0$ the fundamental vector fields 
are $X_0=0$ and $X_1=y\pd{}y$. The Lie system associated to 
$\dot g\, g^{-1}=-b_0\, a_0-b_1\, a_1$ is  $\dot y=b_1y$
and according to the result of the preceding Theorem, 
once we know  a  solution of this last homogeneous linear equation, we 
can carry out the reduction procedure. More explicitly,  when
we know a solution of  $\dot y=b_1y$, the change of variable $x=y\,\zeta$
will simplify the equation  $\dot x=b_0+b_1\, x$ to one on the subgroup
$H_0$, $\dot \zeta=b_0\, y^{-1}$.

If we consider instead the Lie subgroup 
$H_1=\{(0,a_1)\mid a_1\in {\mathbb R}-\{0\}\}$, 
then the elements of the one-dimensional homogeneous space ${\cal A}_1/H_1$
can be characterized by $z=x_0$.
The expression of the fundamental vector fields in this coordinate system 
are $X_0=\pd{}{z}$ and $X_1=z\pd{}z$. Then, as soon as we know a solution 
of $\dot z=b_1z+b_0$,
namely a particular solution of the inhomogeneous equation, 
we can reduce the problem of finding the general solution of $\dot x=b_0+b_1\, x$
to solving an equation on $H_1$, which is a homogeneous linear equation. 
This procedure corresponds to the change of variables
 $x=z+\zeta$, which leads to the reduced equation $\dot \zeta=b_1\, \zeta$.

Therefore the two methods usually found in textbooks for solving the 
inhomogeneous linear differential equation appear here as particular
 cases of a more general methodology for reduction of differential equation
 systems to simpler ones.  

In this way, the last method can be generalized  
when one considers an inhomogeneous linear system, whose
associated group is the corresponding affine group.
Given a particular solution, the problem is reduced to another one 
on its stabilizer,
i.e. the group $GL(n,{\mathbb R})$, 
or, in other words, to a homogeneous linear system.

As another example, we consider 
the Riccati equation, which has been shown to be an
example of Lie system, corresponding to the left action (\ref{action_SL2}) of 
$SL(2,{\mathbb R})$ on the (compactified) real line $\overline{\mathbb{R }}$
by homographies, see e.g., \cite{Win83,CarRam}. The action of the group
of automorphisms of the principal bundle translates to the 
space $\overline{\mathbb{R}}$ into
an action of the group of curves in $SL(2,{\mathbb R})$ on 
the set of Riccati equations. This action was used 
in \cite{CarRam} for studying the
integrability properties of the Riccati equation.
The stabilizer of the point at the infinity is the affine
group in one dimension ${\cal A}_1$. 
Therefore, if we know a particular solution, $x_1$, 
of the Riccati equation, the problem reduces to one on 
${\cal A}_1$, i.e. a inhomogeneous linear equation, 
by means of the well-known change of variable $x=x_1+z$. 
Had we chosen the origin $0\in \overline{\mathbb{R }}$ as the initial point, 
the stabilizer (isomorphic to  ${\cal A}_1$) 
would be  generated by dilations and cotranslations. 
This corresponds to a 
new reduction of the Riccati equation by means of the change of variable
$$ 
x'=\frac{x}{-\frac{x}{x_1}+1}=\frac{x\,x_1}{x_1-x}\,,
$$
which transforms $\dot x=c_0(t)+c_1(t)\,x+c_2(t)\,x^2$ 
into 
$$
\dot x'=\frac{d x'}{dt}=\left(\frac{2\,c_0(t)}{x_1}+c_1(t)\right) x'+c_0(t)\ ,
$$
with associated group ${\cal A}_1$. For more details, see \cite{CarRam}
and \cite{CarRamGra}.

 Now, suppose we know not only one but two different particular solutions,
 $x_{(1)}$ and $x_{(2)}$, of a Lie system in a homogeneous space. They will 
be determined by the choice of initial conditions which provide different
 presentations of the homogeneous space as $G/H_1$ and $G/H_2$, respectively,
 where $H_i$ is the stability subgroup of $x_{(i)}(0)$. Using the result
 of Theorem 2, with $g_1(t)$ being a lifting to $G$ of both curves  
$x_{(1)}(t)$
and  $x_{(2)}(t)$, we will get an equation like the one in the Theorem but
 where the right hand side will be in the intersection $T_eH_1\cap T_eH_2$, 
and therefore the Lie system is reduced to one on the subgroup $H_1\cap H_2$.
The example of the Riccati equation was explicitly considered
 in \cite{CarRam}, where it was also shown that the knowledge of a third 
solution reduces the 
problem to a trivial equation  $\dot x=0$, and therefore giving rise in 
this way to 
the superposition function (\ref{sfRe}). 

\section{Some applications in Classical and Quantum Mechanics}

Non-autonomous linear systems and  Riccati equations are examples 
of Lie systems that  appear very often
in Physics. For instance, linear systems appear in the time evolution of
time-dependent harmonic oscillators and the latter is a condition for
the super-potential $W$ in the
 factorization of a typical quantum Hamiltonian 
$H=-d^2/{dx^2}+V(x)$ as  $H-\epsilon=(-d/{dx} +W)( d/{dx} +W)$, 
where $\epsilon$ is a constant (see, e.g., \cite{CarMarPerRan,CarRamdos,CarRamcuat}),
and it plays a relevant r\^ole in the search for the so-called Shape Invariant
potentials (see \cite{CGM01,CarRamdos,Gen,CarRamtres,IH}). 
As we have pointed out in preceding sections,
the Riccati equation may appear each time that the group $SL(2,\R)$ plays a r\^ole,
and because of the isomorphism of the Lie algebras of 
$SL(2,\R)$ and the linear symplectic group in two dimensions, 
it will be useful in the linear approximation of symplectic 
transformations  and the theory of aberrations in 
optics \cite{CarNas}.

However, the Riccati equation is particularly important because it appears 
as a consequence of Lie reduction theory when taking into account 
that dilations are symmetries of linear second order differential equations
\cite{CarMarNas}.
 Actually, 
the homogeneous linear second-order differential equation
\begin{equation}
\frac{d^2 z}{dx^2}+b(x)\frac{d z}{dx}+c(x)z=0\,,     \label{lsode}   
\end{equation}
admits as an infinitesimal symmetry the vector field 
$X=z\,\partial/{\partial z}$ generating dilations 
 in the variable $z$, which is defined for $z\neq 0$.
According to Lie theory  we
should  change the coordinate $z$ to a new one, $u=\varphi(z)$, such
that  $X={\partial}/{\partial u}$.
This change is determined by the
equation $Xu=1$, which leads to $u=\log |z|$, i.e. $|z|=e^{u}$. In both cases
of regions with $z>0$ or $z<0$ we have 
$$
\frac{dz}{dx}=z\frac{du}{dx}\,,\quad\quad\mbox{and}
\quad\quad\frac{d^2 z}{dx^2}=z\bigg(\frac{du}{dx}\bigg)^2+z\,\frac{d^2 u}{dx^2}\,,
$$
so the equation (\ref{lsode}) becomes
$$
\frac{d^2 u}{dx^2}+b(x)\frac{d u}{dx}+\left(\frac{du}{dx}\right)^2+c(x)=0\,,
$$
and the order can be lowered 
 by introducing the new 
variable $w={du}/{dx}$. We arrive to the following Riccati equation for $w$ 
\begin{equation}
\frac{dw}{dx}=-w^2-b(x)w-c(x)\,.\label{redric}
\end{equation}

Notice that $w=z^{-1}dz/dx$, and that this relation together with (\ref{redric})
 is equivalent
 to the original second order equation. In the particular case 
of the one-dimensional
time-independent Schr\"odinger equation
$$
-\frac{d^2 \phi}{dx^2}+(V(x)-\epsilon)\phi=0\,, 
$$
the reduced Riccati equation for $W= \phi^{-1}{d\phi}/{dx}$ is 
\begin{equation}
W^\prime=-W^2+(V(x)-\epsilon)\,,\label{RaSe}
\end{equation}
which is the equation that $W$ must satisfy in the previously mentioned 
factorization of $H=-d^2/{dx^2}+V(x)$.

Equations of type (\ref{RaSe}) are
particular cases of Riccati equations. We have shown in a preceding 
section that it is possible to  act
with the group of curves in $SL(2,\R)$ on the set of Riccati equations. That means
that a given Riccati equation can be transformed in other related equations (for more
explicit  details see, e.g., \cite{CarRam}). Therefore, when using  curves in 
$SL(2,\R)$ preserving the form of a given equation like (\ref{RaSe}), 
we are transforming the spectral problem for a given Hamiltonian 
into that of another one. This method is carefully explained in \cite{CFR}, 
where explicit examples of the usefulness of the theory are given.  

In a typical problem of Classical Mechanics we are dealing with
Hamiltonian  vector fields in a symplectic manifold $(M,\Omega)$, and then
  we should consider the case in which 
the vector fields arising in the expression of the $t$-dependent vector field describing 
a Lie system, are Hamiltonian vector fields closing on a real 
finite-dimensional Lie algebra. These vector fields correspond to a 
symplectic action of the group $G$ on the symplectic manifold $(M,\Omega)$. 
The Hamiltonian functions of such vector fields,
defined by $i(X_\alpha)\Omega=-dh_\alpha$, however, do not close on the same Lie 
algebra when the Poisson bracket is considered, but we can only say that 
$$d\left(\{h_\alpha,h_\beta\}-h_{[X_\alpha,X_\beta]}\right)=0\ ,
$$
and therefore they span a Lie algebra extension of the original one.
   
The situation in Quantum Mechanics is quite similar. It is well-known that 
the separable complex Hilbert space of states $\cal H$ can be seen as a real manifold admitting a global chart \cite{BCG}. The Abelian translation group allows us to 
identify the tangent space $T_\phi\cal H$ at any point $\phi\in
\cal H$  with  $\cal H$ itself,  the isomorphism being obtained by
associating with $\psi\in\cal H$ the vector $\dot{\psi}\in
T_\phi\cal H$ given 
by: 
$$ 
\dot{\psi}f(\phi) := \left(\frac d{dt}f(\phi+t\psi)\right)_{|t=0}\ ,
$$ 
for any $f\in C^\infty(\cal H)$.

The symplectic 2-form $\W$ is given by  
$$ 
\Omega_{\phi}(\dot{\psi},\dot{\psi'}) = 2\,\ima \<\psi|\psi'>\ , 
$$
with $\<\cdot|\cdot>$ denoting the Hilbert inner product on $\cal H$.

Through the  identification of $\cal H$ with $T_\phi\cal H$ a continuous
vector field is just a continuous
map $A\colon \cal H\to \cal H$; therefore 
a linear operator $A$ on $\cal H$ is a special kind of vector field.

Given a smooth function $a\colon \cal H\to\R$, its differential
$da_\phi $ at $\phi\in \cal H$ is an element of the (real) dual ${\cal H}'$ given 
by: 
$$
\brakt{da_\phi}\psi:=\left({\frac{d}{dt}}a(\phi+t\psi)\right)_{|t=0}\ .
$$

Now, as it was pointed out in \cite{BCG} the skew-Hermitian linear operators
in $\cal H$ define Hamiltonian vector fields, the Hamiltonian function of $-i\, A$
for a self-adjoint operator $A$ being $a(\phi)=\frac 12 \<\phi,A\phi>$. The 
Schr\"odinger equation plays the r\^ole of Hamilton equations because 
it determines the integral curves of the vector field $-i\,H$.

Now, Lie system theory applies to the case in which a $t$-dependent 
Hamiltonian can be written as a linear combination with $t$-dependent 
coefficients of Hamiltonians $H_i$ closing on, under the commutator bracket, 
a real finite-dimensional Lie algebra. The remarkable point, however, is that 
this Lie algebra does not necessarily coincide with the corresponding classical
 one, but it is a Lie algebra  extension. An example will be given in next section.

\section{An example: classical and quantum time-dependent linear potential}

The linear potential model, with many applications in physics, has recently been studied by
Guedes \cite{Gu01}. We can use this problem in order to 
 illustrate the possible applications
 of the theory by means of a simple example. Let us consider the classical system 
described by a classical Hamiltonian
 $$H_c=\frac{ p^2}{2m}+f(t)\, x\ ,
 $$
 and the corresponding quantum Hamiltonian
$$H_q=\frac{P^2}{2m}+f(t)\, X\ ,
$$
describing, for instance when $f(t)=q\, E_0+q\,E\,\cos\omega t $, the motion of a particle
of electric charge $q$ and
mass $m$ driven by a monochromatic electric field. $E_0$ is the strength of
 the constant confining electric field and $E$ that of the time-dependent
 electric field
that drives the system with a frequency $\omega/2\pi$.
Instead of using the Lewis and Riesenfeld invariant method \cite{LR69} as
it was done in \cite{Gu01}, we will study simultaneously the classical and the quantum problem
by reduction of both problems to  similar equations and using the Wei--Norman 
method to
solve such an equation. As it will be shown, the only difference is that the
 Lie algebra arising in the quantum problem is not the same as in the classical one, 
but a central extension.  

The classical Hamilton equations of motion are
\begin{eqnarray}{\dot x}&=&\frac pm\,,\cr
 {\dot p}&=&-f(t)\,,\end{eqnarray}
and therefore,
the motion is given  by
\begin{eqnarray}
x(t)&=&x_0+\frac{p_0\, t}m-\frac 1m\int_0^tdt'\,\int_0^{t'}f(t'')\, dt''\ ,\\
p(t)&=&p_0-\int_0^tf(t')\,dt'\label{emls}
\end{eqnarray}

The $t$-dependent vector field describing the time evolution  is
$$X=\frac pm\,\pd{}x-f(t)\,\pd{}p\ .
$$

This vector field can be written as a linear combination
$$X=\frac 1m\, X_1-f(t)\, X_2\ ,
$$
with
$$
X_1=p\, \pd{}x\ ,\qquad X_2=\pd{}p\ ,
$$
being two vector fields closing a three dimensional Lie algebra with 
$$X_3=\pd{}x\ ,
$$
isomorphic to the Heisenberg algebra, namely,
\begin{equation}
[X_1,X_2]=-X_3\ , \qquad [X_1,X_3]=0\ ,\qquad [X_2,X_3]=0\ .\label{Xla}
\end{equation}

The flow of these vector fields is given, respectively, by
\begin{eqnarray}
&&\phi_1(t,(x_0,p_0))=(x_0+p_0\, t,p_0)\,,      \nonumber\\
&&\phi_2(t,(x_0,p_0))=(x_0,p_0+t)\,,            \nonumber\\
&&\phi_3(t,(x_0,p_0))=(x_0+t,p_0)\,.            \nonumber
\end{eqnarray}

In other words, this corresponds to the action of the Lie group of 
upper triangular $3\times 3$ matrices 
on $\R^2$\ ,
$$
\matriz{c}{\bar x\\ \bar p\\1}=\matriz{ccc}{1&a_1&a_3\\0&1&a_2\\0&0&1}
\matriz{c}{x\\ p\\1}\ .
$$

It is to be remarked that the three vector fields $X_1$, $X_2$ and $X_3$ are 
Hamiltonian vector fields with respect to the usual symplectic structure, 
$\Omega=dx\wedge dp$, the corresponding Hamiltonian functions $h_i$ such that 
$i(X_i)\Omega=-dh_i$ being 
$$h_1=-\frac {p^2}2\,,\qquad h_2=x\,,\qquad h_3=-p\ ,$$
therefore
\begin{equation}\{h_1,h_2\}=-h_3\,, \quad \{h_1,h_3\}=0\,, \quad \{h_2,h_3\}=-1\,, 
\label{pbalg}
\end{equation}
which close on a four-dimensional Lie algebra with $h_4=1$, that is,
a central  extension of that given by (\ref{Xla}). Let $\{a_1,\,a_2,\,a_3\}$ be a basis
of the Lie algebra with non-vanishing defining relations $[a_1,a_2]=-a_3$. 
Then, the corresponding equation in the group (\ref{eqingr}) becomes in this case 
$$
\dot g\, g^{-1}=-\frac 1 m a_1+f(t)\, a_2\ .
$$
Now, choosing the factorization
$g= \exp(-u_3\, a_3)\,\exp(-u_2\, a_2)\,\exp(-u_1\, a_1)$ and using 
the Wei--Norman formula (\ref{eq_met_WN})
we will arrive to the system of differential equations
$$
\dot u_1=\frac 1m\ ,\qquad \dot u_2=-f(t)\ ,\qquad
\dot u_3-\dot u_1\, u_2=0\ ,
$$
together with the initial conditions
$$u_1(0)=u_2(0)=u_3(0)=0\ ,
$$
with solution 
$$
u_1=\frac tm\ ,\qquad u_2=-\int_0^tf(t')\, dt'\ ,
\qquad u_3=-\frac 1m \int_0^t dt'\int_0^{t'}f(t'')\, dt''\ .
$$

Therefore the motion will be given by
$$
\matriz{c}{x\\p\\1}= \matriz{ccc}{1&\frac tm&-\frac 1m\int_0^t dt'\int_0^{t'}f(t'')\,
 dt''\\0&1&-\int_0^tf(t')\, dt'\\0&0&1}\matriz{c}{x_0\\p_0\\1}\ ,
$$
which reproduces (\ref{emls}). So, obviously we will recover the constant of
motion given in \cite {Gu01}, $I_1=p(t)+\int_0^tf(t')\, dt'$, together with
the other one $I_2=x(t)-\frac 1m\left(p(t)+\int_0^tf(t')\, dt'\right) t
+\frac 1m\int_0^tdt'\int_0^{t'}f(t'')\, dt''$.

As far as the quantum problem is concerned, also studied in a very recent 
paper \cite{Ba01}, notice that the quantum Hamiltonian
$H_q$ may be written as a sum
$$
H_q=\frac 1 m \, H_1-f(t)\, H_2\ ,
$$
with
$$
H_1=\frac {P^2}2\ ,\qquad  H_2=-X\ ,
$$
and $-i\, H_1$ and $-i\, H_2$  close on a four-dimensional Lie algebra 
with $-i\, H_3=-i\,P$, and $-i H_4=i\,I$, isomorphic to that of (\ref{pbalg}),
which is an extension of the Heisenberg 
Lie algebra  (\ref{Xla}),
$$
[-i\,H_1,-i\, H_2]=-i\,H_3\,,\ [-i\,H_1,-i\, H_3]=0\,,\ [-i\,H_2,-i\, H_3]=-i\,H_4\,.
$$

The Schr\"odinger equation given by the Hamiltonian $H_q$ is like that of 
a Lie system. Note that this Hamiltonian is time--dependent and that such systems 
are seldom studied, because it is generally difficult to find the time evolution 
of such systems. However, this system is a Lie system and therefore we can find 
the time-evolution operator by applying the reduction of the problem to an
equation on the Lie group and using the Wei--Norman  method.

Let $\{a_1,\,a_2,\,a_3,\,a_4\}$ be a basis of the Lie algebra with 
non-vanishing defining relations $[a_1,a_2]=a_3$ and $[a_2,a_3]=a_4$.
The  equation  (\ref{eqingr}) in the group to be 
considered is now
$$
\dot g\, g^{-1}=-\frac 1 m\, a_1+f(t)\, a_2\ .
$$
Using  the factorization 
$g=\exp(-u_4\, a_4)\exp(-u_3\, a_3)\,\exp(-u_2\, a_2)\,\exp(-u_1\, a_1)$ 
the Wei--Norman method provides the following equations:
\begin{eqnarray}
&\dot u_1=\frac 1m\,, \quad\quad &\dot u_2=-f(t)\,,                     \nonumber\\
&\quad\quad\quad\dot u_3+u_2\, \dot u_1=0\,, 
\quad\quad &\dot u_4+u_3\,\dot u_2-\frac 12\, u_2^2 \,\dot u_1=0\,,       \nonumber
\end{eqnarray} 
and written in normal form 
\begin{eqnarray}
&\dot u_1=\frac 1m\,, \quad\quad &\dot u_2=-f(t)\,,                     \nonumber\\
&\quad\quad\dot u_3=-\frac 1m\,u_2\,, \quad\quad &\dot u_4=f(t)\, u_3+\frac 1{2m} u_2^2\,,
                                                                \nonumber
\end{eqnarray}
together with the initial conditions 
$u_1(0)=u_2(0)=u_3(0)=u_4(0)=0$, whose solution is 
$$
u_1(t)=\frac tm\,,\quad u_2(t)=-\int_0^tf(t')\, dt'\,,\quad 
u_3(t)=\frac 1m\int_0^tdt'\int_0^{t'}f(t'')\, dt''\,,
$$
and 
$$
u_4=\frac 1m\int_0^t dt' f(t')\int_0^{t'}dt''\int_0^{t''}f(t''')\, dt'''
+\frac 1{2m}\int_0^t dt'\, \left(\int_0^{t'}dt''f(t'')\right)^2\ .
$$
These functions provide the explicit form of the time-evolution operator:
$$
U(t,0)=\exp(-i u_4(t))\exp(i u_3(t) P)\exp(-i u_2(t) X)\exp(i u_1(t) P^2/2)\ .
$$

Notwithstanding, in order to find the expression of the wave-function in a
simple way, it is advantageous to use the factorization 
$$
g=\exp(-v_4\, a_4)\exp(-v_2\, a_2)\,\exp(-v_3\, a_3)\,\exp(-v_1\, a_1)\,.
$$
In such a case, the Wei--Norman method gives the system
\begin{eqnarray}
&\dot v_1=\frac 1m\,, \quad\quad &\dot v_2=-f(t)\,,                     \nonumber\\
&\quad\quad\dot v_3=-\frac 1m\,v_2\,, \quad\quad &\dot v_4=-\frac 1{2m} v_2^2\,,
                                                                \nonumber
\end{eqnarray}
jointly with the initial conditions $v_1(0)=v_2(0)=v_3(0)=v_4(0)=0$. The solution is
\begin{eqnarray}
&&v_1(t)=\frac tm\,,\quad v_2(t)=-\int_0^t dt'\,f(t')\,,\quad           \label{vs12}\\ 
&&v_3(t)=\frac 1 m \int_0^t dt'\int_0^{t'} dt'' f(t'')\,,               \label{vs3}\\ 
&&v_4(t)=-\frac 1{2m}\int_0^t dt'\left(\int_0^{t'} dt'' f(t'')\right)^2\,. \label{vs4} 
\end{eqnarray}
Then, applying the evolution operator onto the initial wave-function $\psi(p,0)$, which
is assumed to be written in momentum representation, we have
\begin{eqnarray}
\psi(p,t)&=&U(t,0)\psi(p,0)                                                     \nonumber\\
&=&\exp(-i v_4(t))\exp(-i v_2(t) X)\exp(i v_3(t) P)\exp(i v_1(t) P^2/2)\psi(p,0)               \nonumber\\
&=&\exp(-i v_4(t))\exp(-i v_2(t) X)e^{i(v_3(t) p+v_1(t) p^2/2)}\psi(p,0)                             \nonumber\\
&=&\exp(-i v_4(t))e^{i(v_3(t) (p+v_2(t))+v_1(t) (p+v_2(t))^2/2)}\psi(p+v_2(t),0)\,,                           \nonumber
\end{eqnarray}
where the functions $v_i(t)$ are given by (\ref{vs12}), (\ref{vs3}) and (\ref{vs4}),
respectively.


\section*{Acknowledgments}

A.R. thanks the Spanish Ministerio de Ciencia y Tecnolog\'{\i}a for a FPI grant.
Partial support of the Spanish DGES is also acknowledged.

\end{document}